%% file: main.tex
\documentclass[letterpaper]{article}
\usepackage{aaai}
\usepackage{times}
\usepackage{helvet}
\usepackage{courier}
\usepackage{graphicx}
\usepackage{caption}
\usepackage{subcaption}
\usepackage{color,url,amsmath}
\usepackage{verbatim}
\usepackage{subcaption}
\usepackage{adjustbox}
\usepackage{hyperref}
\usepackage[table]{xcolor}

\begin{document}
\title{ The VGLC: The Video Game Level Corpus}
%\title{Super Metric Bros.: Towards Computational Metrics \\ that can Accurately Predict Human Ratings \\ {\color{red} [Adam: This title seems way too strong.  It should at least mention platformer levels and more specifically probably Mario -it also feels a bit dry to me, but I like puns in my titles] [Levi How about now? :)]}}
%\title{Super Metric Bros.: Understanding Platformer Level Designs \\ with Interpretable Metrics}
\author{Adam James Summerville\textsuperscript{1}, Sam Snodgrass\textsuperscript{2},  Michael Mateas\textsuperscript{1}, and Santiago Onta\~{n}\'{o}n\textsuperscript{2}\\
\textsuperscript{1}University of California, Santa Cruz \\
1156 High Street \\
Santa Cruz, CA 95066\\ 
\textsuperscript{2}Drexel University \\
3141 Chestnut Street \\
Philadelphia, PA 19104 \\
\\ 
asummerv@ucsc.edu, sps74@drexel.edu, michaelm@soe.ucsc.edu, santi@cs.drexel.ed
}
% asummerv@ucsc.edu, michaelm@soe.ucsc.edu  }
\maketitle
   \section*{ABSTRACT}
   Levels are a key component of many different video games, and a large body of work has been produced on how to procedurally generate game levels.  Recently, Machine Learning techniques have been applied to video game level generation towards the purpose of automatically generating levels that have the properties of the training corpus.  Towards that end we have made available a corpora of video game levels in an easy to parse format ideal for different machine learning and other game AI research purposes.
\section*{Keywords}
Video Games, Level Design, Procedural Content Generation, Machine Learning, Corpus

\section*{INTRODUCTION}

For many different video games, levels are one of the critical pieces of game content.  They represent the virtual space wherein the majority of player interaction occurs.  As such, they represent a very attractive target for Procedural Content Generation (PCG), i.e. the creation of artefacts via an algorithm.  Most PCG level creation has been accomplished via human-authored rules from early computer games such as \textit{Rogue} \nocite{ROGUE} up through modern games such as \textit{No Man's Sky}. % \nocite{NOMANSSKY}. % commented, it since, otherwise, there was a space between the name and the dot 
A large body of academic work has been performed in this field utilizing classical AI techniques such as constraint satisfaction (\cite{TANAGRA}), Answer Set Programming (\cite{REFRACTION}), Evolutionary Algorithms (\cite{SORENSON}), and others.  More recently, statistical AI, i.e. Machine Learning (ML), techniques have been used such as Bayes Nets (\cite{BAYESZELDA}), Markov Chains (\cite{SNODGRASS}, \cite{MCMCTS}, \cite{DAHLSKOG}), clustering (\cite{GUZDIAL}), non-negative matrix factorization (\cite{SHAKER}), PCA (\cite{BAYESZELDA}), and others.  While a large number of different ML techniques have been used they all have one thing in common, they require a training corpus.  

Towards that end we have assembled a corpus consisting of 428 levels from 12 games\footnote{The corpus can be found at: \href{https://github.com/TheVGLC/TheVGLC}{https://github.com/TheVGLC/TheVGLC}}.  These levels exist as parseable text files, along with the corresponding image representation of the level, ideal for machine consumption for either ML PCG or other game AI applications. In the following sections we will first discuss the games in the corpus and the details of the levels included; we will then show a small subset of work that has been done with these levels; and finally we will discuss ways these levels could be used in the future.

\begin{table*}[ht]
\begin{adjustbox}{width=1.0\textwidth,center}
\begin{tabular}{ | l | c | c | c | c |}
\hline
  Game & Levels & Minimum Sizes & Median Sizes & Maximum Sizes \\  
\hline
  \textit{Super Mario Bros.} & 20 \textbf{Tile} & 150  x 14 & 188 x 14 & 374 x 14 \\
  \textit{Super Mario Bros. 2} & 25 \textbf{Tile} & 161  x 12 & 194 x 15 & 364 x 16 \\
  \textit{Super Mario Land} \nocite{SML} & 9 \textbf{Tile} & 261  x 15 & 294 x 16 & 441 x 16 \\
  \textit{Super Mario Kart} \nocite{MARIO_KART}& 7 \textbf{Tile} & 128  x 128 & 128 x 128 & 128 x 128 \\
  \textit{Kid Icarus} \nocite{KID_ICARUS} & 6 \textbf{Tile} & 16  x 159 & 16 x 205 & 16 x 281 \\
  \textit{Lode Runner} & 150 \textbf{Tile} & 33  x 22 & 33 x 22 & 33 x 22 \\
  \textit{Rainbow Islands} \nocite{RAINBOW_ISLANDS} & 28 \textbf{Tile} & 32  x 83 & 33 x 165 & 33 x 252 \\
  \textit{Doom} \nocite{DOOM} & 36 \textbf{Tile} & 36  x 69 & 121 x 117 & 225 x 213 \\
  \textit{Doom 2} \nocite{DOOM2} & 32 \textbf{Tile} & 54  x 60 & 131 x 123 & 231 x 274 \\
  \textit{The Legend of Zelda} \nocite{LOZ} & 9 \textbf{Tile} & 67  x 32 & 89 x 80 & 89 x 128 \\
  \textit{The Legend of Zelda} & 18 \textbf{Graph} & $ \mid V \mid$  = 12  $\mid E \mid$  = 22 $\Delta$(G) = 3& $ \mid V \mid $ = 27 $ \mid E \mid $ = 58 $\Delta$(G) = 4&  $ \mid V \mid $ = 66 $ \mid E \mid $ = 161 $\Delta$(G) = 6\\
 \begin{tabular}{@{}c@{}} \textit{The Legend of Zelda:} \\ \textit{A Link to the Past} \end{tabular} \nocite{LTTP} & 12 \textbf{Graph} &  $\mid V \mid$  = 14  $\mid E \mid$  = 31 $\Delta$(G) = 2&  $\mid V \mid$  = 34 $ \mid E \mid$  = 76 $\Delta$(G) = 4&   $\mid V \mid$  = 65  $\mid E \mid$  = 125 $\Delta$(G) = 8\\
 \begin{tabular}{@{}c@{}} \textit{The Legend of Zelda:} \\ \textit{Link's Awakening } \end{tabular}  \nocite{LINKS_AWAKENING} & 8 \textbf{Graph} &  $\mid V \mid$  = 21  $\mid E \mid$  = 43 $\Delta$(G) = 3&  $\mid V \mid$  = 43  $\mid E \mid$  = 98 $\Delta$(G) = 5&   $\mid V \mid$  = 59  $\mid E \mid$  = 136 $\Delta$(G) = 8\\
  \textit{Doom} & 36 \textbf{Vector} & 122 lines, 53 objects  & 956 lines, 251 objects& 1764 lines, 463 objects \\
  \textit{Doom 2} & 32 \textbf{Vector} & 93 lines, 69 objects  & 774 lines, 253 objects& 1690 lines, 509 objects\\
  \hline
\end{tabular}
\end{adjustbox}
\caption{The games included in the corpus as of publication}
\end{table*}
\section*{RELATED WORK}

The collection of resources for the creation of a shared corpus is commonplace in fields such as Natural Language Processing (NLP). A large number of corpora exist in the NLP community, but the two most noteworthy are  COCA: "the Corpus of Contemporary American English" (\cite{COCA}) and the Wall Street Journal Corpus (\cite{WSJ}).  Similarly, many corpora are available in the machine learning community for a variety of different tasks including image classification (\cite{CIFAR}), hand-writing recognition (\cite{MNIST}), and speech-recognition (\cite{SWITCHBOARD}).  

Machine learning based approaches for the procedural generation of video game levels have relied on a couple of different data sources: synthetic data, annotated images, and video.  While a commonplace activity in many machine learning communities, the only known videogame level PCG to use synthetic data is the work of Shaker and Abou-Zleikha (\cite{SHAKER}) which used data generated by other generation systems.  A number of works have used annotated level images (either annotated by human hand or automatically via image-processing), mostly of \textit{Super Mario Bros.} \nocite{SUPER_MARIO_BROS} (\cite{MCMCTS}, \cite{SMBRNN}, \cite{SNODGRASS}, \cite{SNODGRASS2}, \cite{DAHLSKOG}, \cite{HOOVER}, \cite{GRAPH_GRAMMAR_SMB}) but occasionally other games such as those from \textit{The Legend of Zelda} series (\cite{BAYESZELDA}, \cite{SAMPLING_HYRULE}) or \textit{Lode Runner} \nocite{LODE_RUNNER} (\cite{SNODGRASS2}).  The work of \cite{GUZDIAL} used Long Play videos of \textit{Super Mario Bros.} gathered from Youtube to learn the spatial relationships of different sprite groupings.

\section*{DATASETS}

The levels from 12 games are present in the corpus as of the publication of this paper.  There are three annotation formats: \textbf{Tile}, \textbf{Graph}, and \textbf{Vector}.  A breakdown of the games can be seen in table 1. Most of the games in the corpus are 2D tile-based sidescrolling games which are all annotated with the \textbf{Tile} format, as are the levels from Doom, Doom 2, Mario Kart, and the first quest of \textit{The Legend of Zelda}.  The levels from \textit{The Legend of Zelda} series which have a room based level structure are also annotated in the \textbf{Graph} format, and the levels from the \textit{Doom} series can also be found in the \textbf{Vector} format.

\subsection*{Annotation Formats}
We will now detail the annotation formats.  The \textbf{Tile} format is an intuitive format for tile based games, particularly those where reasoning over the space of the game should be done in a tile grid.  These tile based levels exist as a two dimensional grid of size $w \times h$ with $w$ being the width and $h$ being the height.  Each entry in this grid is annotated by a single character.  Along with the $w \times h$ array for each level there is a JSON file that acts as a legend.  Each tile character is an entry in a dictionary named \textit{tiles} and has an associated array of possible annotation tags, e.g.  for \textit{Super Mario Bros.} the tile character for the \includegraphics{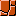} tile has the tags \textbf{[solid, ground]} and the character for the \includegraphics{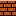} has the tags \textbf{[solid,breakable]}.  The JSON legend file for \textit{Super Mario Bros.} can be seen in figure \ref{fig:SMBLegend}.

\begin{figure*}
\centering
\begingroup
    \fontsize{8pt}{10pt}\selectfont
\begin{verbatim}
{"tiles" : {"X" : ["solid","ground"],
            "S" : ["solid","breakable"],
            "-" : ["passable","empty"],
            "?" : ["solid","question block", "full question block"],
            "Q" : ["solid","question block", "empty question block"],
            "E" : ["enemy","damaging","hazard","moving"],
            "<" : ["solid","top-left pipe","pipe"],
            ">" : ["solid","top-right pipe","pipe"],
            "[" : ["solid","left pipe","pipe"],
            "]" : ["solid","right pipe","pipe"],
            "o" : ["coin","collectable","passable"],
            "B" : ["solid","bullet bill","hazard","enemy"],
            "b" : ["solid","bullet bill"]  } }
\end{verbatim}
\endgroup
\caption{Legend file for Super Mario Bros.}
\label{fig:SMBLegend}
\end{figure*}
\begin{figure*}
\centering

\begin{subfigure}[t]{0.5\textwidth}
\includegraphics[height=1in]{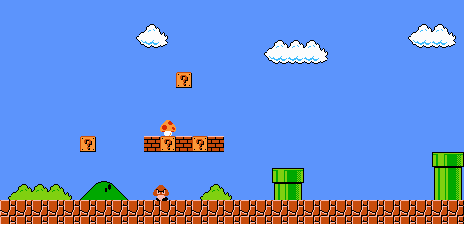}
%\caption{Section of level 1-1 in original sprite version.}
\end{subfigure}
~
\begin{subfigure}[t]{0.32\textwidth}
\centering
\includegraphics[height=1in]{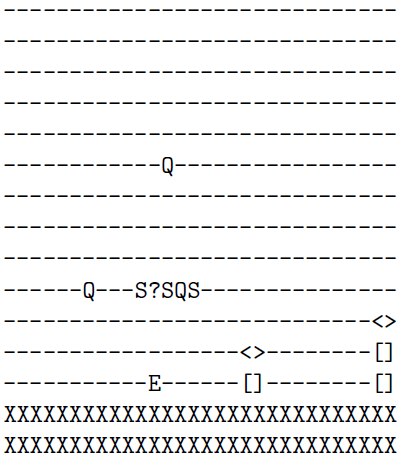}
%\caption{Annotated text file version.}
% * <sam.psnodgrass@gmail.com> 2016-06-22T21:12:57.503Z:
%
% ^.
\end{subfigure}
\caption{A section of level 1-1 (left) and the annotated text file version (right).}
\label{fig:mariosnippet}

\end{figure*}
A section of the annotated version of Level 1-1  of \textit{Super Mario Bros.}  alongside the image it is derived from can can be seen in Figure \ref{fig:mariosnippet}.  It should be noted that multiple different images can be mapped to the same tile character, e.g. all enemies are mapped to the same character, ``E'', and all solid unbreakable tiles (ground, stairs, tree tops, giant mushroom tops) are mapped to the same character, ``X''.

The \textbf{Graph} annotation format is used for games where the high-level topology should be reasoned about at differently than the low-level structure.  Games with discrete room-to-room structures, e.g. \textit{The Legend of Zelda}, or classic adventure games such as \textit{Zork} \nocite{ZORK} and \textit{King's Quest}\nocite{KINGS_QUEST} are well represented by graphs.  The graph format we chose was the DOT language used by Graphviz (\cite{GRAPHVIZ}).  This format was chosen for 3 reasons:
\begin{itemize}
  \setlength\itemsep{0mm}
\item {\bf Easily Parseable} - The format is very easily parsed with nodes being represented by \verb|<|\textit{Node ID}\verb|> [label="<|\textit{Node Label}\verb|>"]| and edges being represented by  \\ \verb|<|\textit{Source ID}\verb|> -> <|\textit{Target ID}\verb|> [label="<|\textit{Edge Label}\verb|>"]|
\item {\bf Easily Visualized} - As part of Graphviz, DOT files can be consumed by the \textit{dot} program to visualize the graphs
\item {\bf Portable} - Because it is a popular, well-documented format it is able to be used by other programs
\end{itemize}

As with the \textbf{Tile} format there is a corresponding JSON file that acts as a legend for each game.  The \textit{Legend of Zelda} legend can be seen below:

\begingroup
    \fontsize{8pt}{10pt}\selectfont
\begin{verbatim}
{"vertices" :{"e" : ["enemy" ],
              "S" : ["switch" ],
              "b" : ["boss" ],
              "k" : ["key"],
              "K" : ["boss key"],
              "I" : ["key item"],
              "p" : ["puzzle"],
              "s" : ["start"],
              "t" : ["triforce"]},
 "edges" : {  "S" : ["switch locked" ],
    		  "b" : ["bombable" ],
              "k" : ["key locked"],
              "K" : ["boss key locked"],
              "I" : ["key item locked"],
              "l" : ["soft locked"],
              "S" : ["switch locked"],
              "s" : ["visible", "impassable"]}}
\end{verbatim}
\endgroup

\begin{figure*}
\centering
\begin{subfigure}[t]{0.4\textwidth}
\resizebox{50mm}{!}{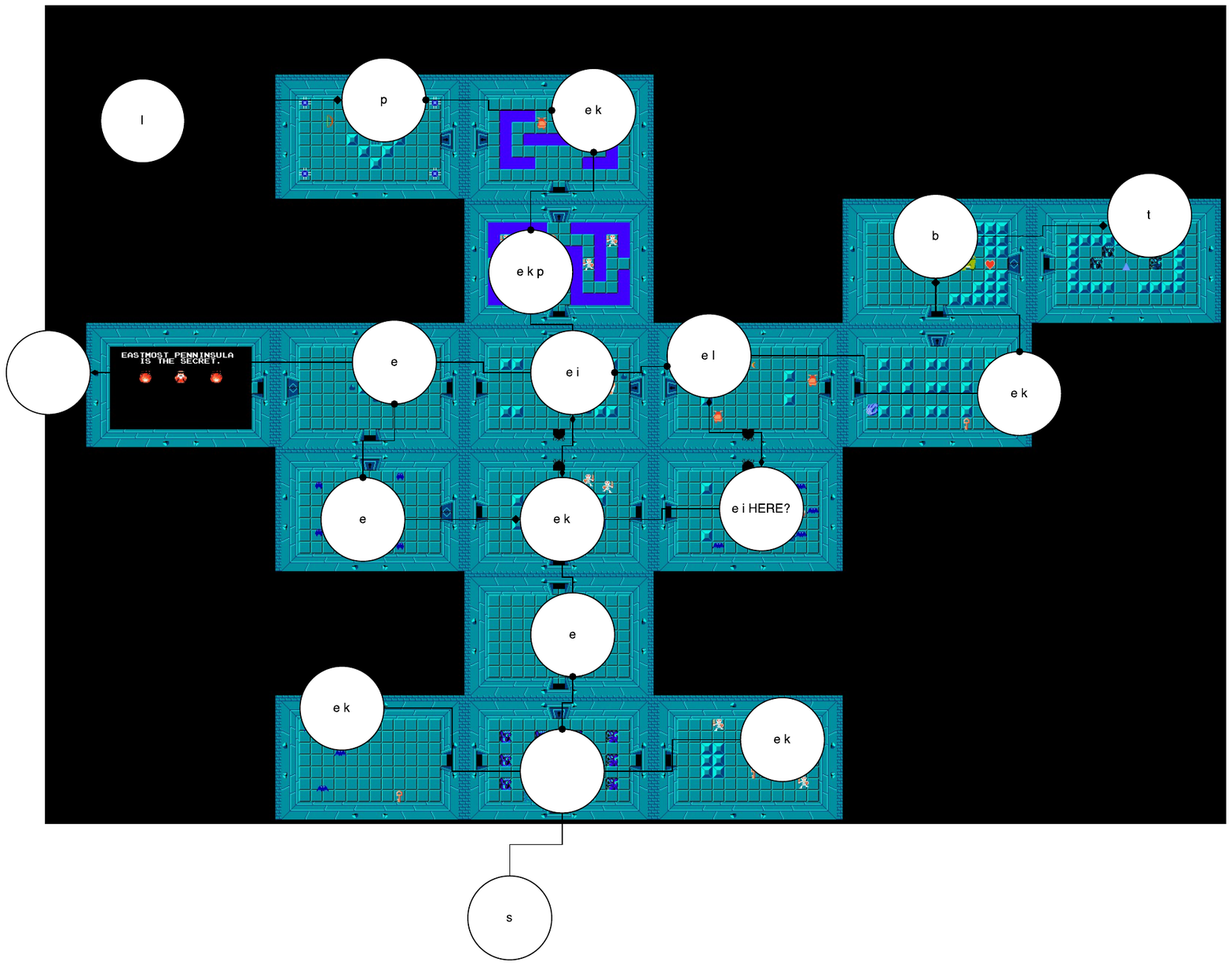}
\end{subfigure}
~
\begin{subfigure}[t]{0.4\textwidth}
\resizebox{50mm}{!}{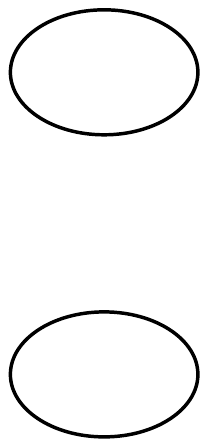}
\end{subfigure}
\caption{The annotated tilemap for Level 1 ``The Eagle'' of \textit{The Legend of Zelda} (LEFT) and the corresponding svg output of the DOT file for it (RIGHT).}
\label{fig:zelda}
\end{figure*}

Additonally, below we have a section of the DOT file for the first dungeon from \textit{The Legend of Zelda}, ``The Eagle.'' The original annotation and the output of the DOT file can be seen in Figure \ref{fig:zelda}:

\begingroup
    \fontsize{8pt}{10pt}\selectfont
\begin{verbatim}
digraph {0 [label=""]
  1 [label="e,I"]
  2 [label="I"]
  ...
  17 [label="e,k"]
  18 [label="p"]
  7 -> 8 [label=""]
  8 -> 7 [label=""]
  ...
  3 -> 13 [label="b"]
  13 -> 3 [label="b"]}
\end{verbatim}
\endgroup

The final annotation format is the \textbf{Vector} format.  For this we chose the Scalable Vector Graphics (SVG) format as it is the most readily readable and viewable vector format.  Games such as those in the \textit{Doom} series (which is ostensibly 3D but gameplay exists solely in a 2D plane) are best thought of as line segments (walls, doors, etc.) and objects (enemies, weapons, decorations).  The SVG format readily handles line segments and discrete objects can be handled as graphical primitives (circles, rectangles, ellipses).  Once more, each game has a corresponding JSON legend, \textit{Doom}'s can be seen in figure \ref{fig:DoomLegend}.

\begin{figure*}
\centering
\resizebox{65mm}{!}{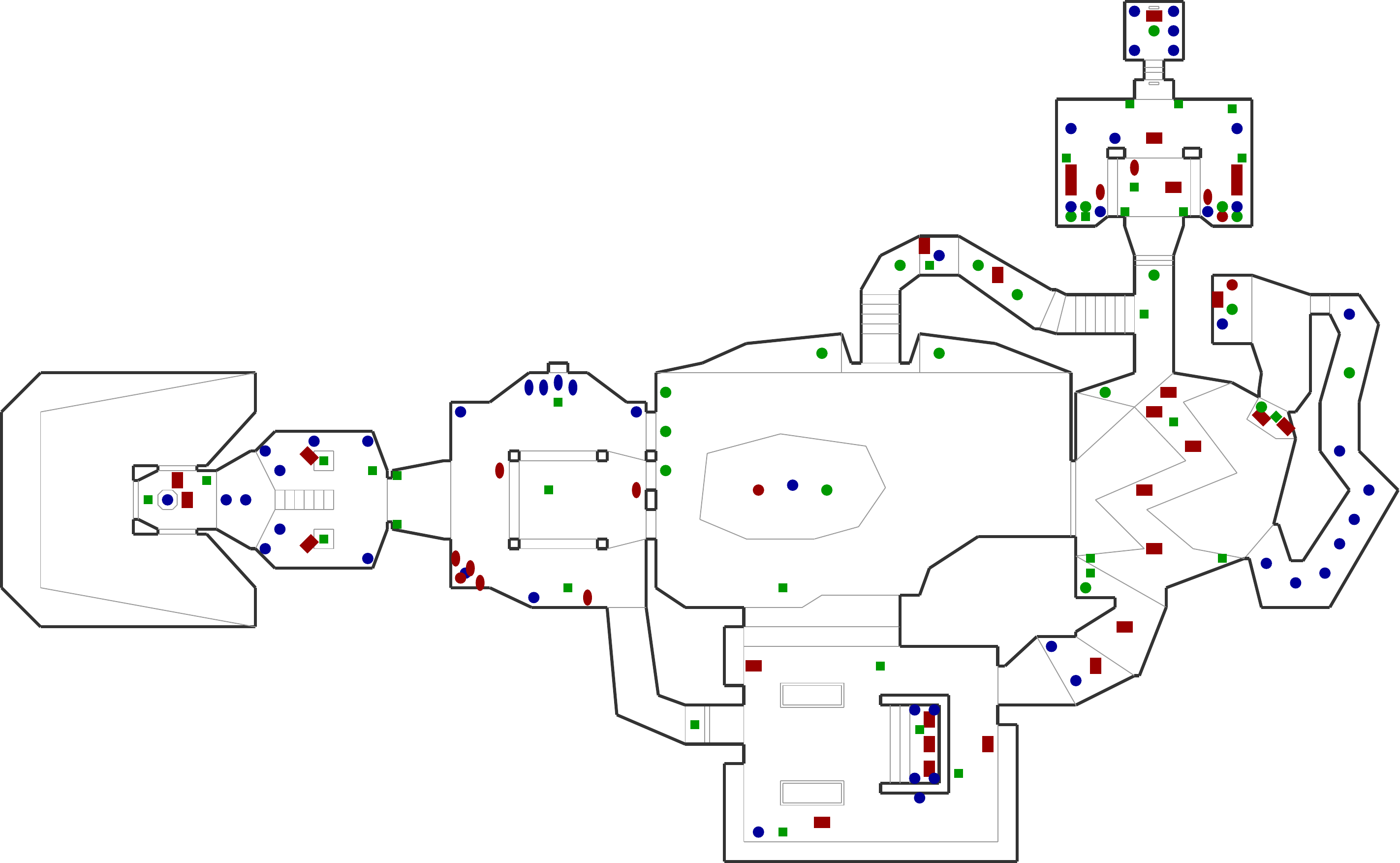}
\caption{The first level of Doom rendered in SVG. Blue circles represent health/armor, green circles represent ammo, red squares represent enemies, red ovals represent explosive barrels, and other shapes correspond to other less common elements such as teleporters, start points, weapons, etc.}
\label{fig:doom}
\end{figure*}
In the legend, the color of the stroke determines the annotation for that line or object primitive.  An SVG representation of \textit{Doom}'s first level can be seen in Figure \ref{fig:doom}.

The annotations explained above are the initial set of annotations we explored. We recognize that there are limitations. For instance, the {\bf Tile} annotation does not have a tile representation for each enemy, and some other special tiles, which results in a loss of information. In the future, we intend to explore more lossless representations, including more expressive tile sets and ontologies for those tile sets, allowing for more complete representations of the levels. Additionally, the original map images are included in the corpus to allow users to create their own annotations.

\begin{figure*}
\begingroup
    \fontsize{8pt}{10pt}\selectfont
\begin{verbatim}
{"element" : { "line" : { "stroke" : {  "#333" : ["solid","wall"],
                                        "#999" : ["floor","walkable","stairs"],
                                        "#d00" : ["door","locked"],
                                        "#00d" : ["teleport","source","activatable"],
                                        "#666" : ["door","walkable","activatable"],
                                        "#0d0" : ["exit","activatable"], } },
               "rect" : { "stroke" : {  "#900" : ["enemy","walkable"],
                                        "#090" : ["decorative","walkable"],
                                        "#009" : ["teleport","walkable","destination"] } },
               "circle" : { "stroke" : {"#900" : ["weapon","walkable"],
                                        "#090" : ["ammo","walkable"],
                                        "#009" : ["health","armor","walkable"] } },
               "ellipse" : {"stroke" : {"#900" : ["explosive barrel","walkable"],
                                        "#090" : ["key","walkable"],
                                        "#009" : ["start","walkable"] } } } }
\end{verbatim}
\endgroup
\caption{Legend for Doom file}
\label{fig:DoomLegend}
\end{figure*}
%In addition to the annotated files, the original images used for the annotations can also be found in the corpus to allow users to create their own annotations, if desired.  
\section*{TOOLS}

Along with the levels, we have also included tools that we have used as part of processing. These include a tile based platformer A$^*$ solver and a parser and rasterizer for Doom WAD files.  The platformer solver takes in a JSON file that acts as a legend (telling it which tiles are solid) as well as the dynamics of the game (what does a jump arc look like) and produces paths through a given level.  The Doom parser and rasterizer take in WAD files (the data format for Doom and Doom-likes) and produces either the SVG or tile based levels found in the corpus.

\section*{POTENTIAL USAGE}

In hope of sparking future research, below are some potential uses of the VGLC.  

\subsection*{Corpora Based Procedural Content Generation}

This is the most obvious use, or at least the most common use for this work so far.  The \textbf{Tile} format is set up in a way where nearly any text generation based approach (Markov chains, recurrent neural networks, etc.) can produce results, but given their grid based nature they are also translatable into a form that can easily be consumed for image based methods (Markov random fields, convolutional neural networks, etc.).  The \textbf{Graph} could be used for graph grammar learning or other graph based approaches (spectral graph analysis, relational learning).  The levels can also be processed such as in the work of Londo\~{n}o and Missura (\cite{GRAPH_GRAMMAR_SMB}) or Summerville and Mateas (\cite{SMBRNN}) where simulated agents are run through levels to determine how a player could actually traverse through the levels.  To our knowledge, the only games to have been used for corpora based generation are \textit{Super Mario Bros.}, \textit{Kid Icarus}, \textit{The Legend of Zelda}, and \textit{Lode Runner} meaning any of the other games are ripe for generation.

\subsection*{Design Analysis}
\addvspace{-0.2\baselineskip}

A large amount of PCG work has relied on assumptions about successful design decisions, but the levels from these games represent actual examples of successful design decisions.  Dahlskog and Togelius (\cite{PATTERNS_PCG}) performed an analysis on 20 of the levels from \textit{Super Mario Bros.} to find design patterns that could be used for PCG research, but even in the \textit{Super Mario Bros.} domain, this corpus contains an additional 34 levels to be analyzed.  Similarly,  Dormans (\cite{DORMANS_ZELDA}) performed an analysis on the mission and physical structure of a level from \textit{The Legend of Zelda: Twilight Princess}, and while none of the 3D Zelda games are represented in this corpus,  
there are 48 levels from the 2D games of the series that could be analyzed.  Beyond those series, the 3 other games could be compared and contrasted with the closer to  saturated \textit{Super Mario Bros.}

\subsection*{Style Transfer}
\addvspace{-0.2\baselineskip}

Along those lines, we do not know of any work that has successfully transferred level design style across different games.  All work has focused on a single game or series, and when work has included multiple games the work has always been partitioned.  Following the work of Gatys et al. (\cite{NEURAL_STYLE}) there has been an interest in applying different artistic styles to images, similarly, we imagine that an interesting avenue for future work would be one that could reimagine levels of one game in the style of another, or a user could sketch the skeleton of a level and in turn generate variants based on different game styles.

\section*{CONCLUSIONS AND FUTURE WORK}
\addvspace{-0.2\baselineskip}

We present this corpus in hopes of helping the community.  Each group of researchers that have used corpora based machine learning approaches have needed to reinvent this, admittedly, not altogether exciting wheel, which is why we expect this work to be adopted by the community so that focus can be placed on more exciting and innovative work. The VGLC is already available online (\href{https://github.com/TheVGLC/TheVGLC}{https://github.com/TheVGLC/TheVGLC}), ready to use.  Furthermore, we encourage researchers to contribute additional games and tools to this corpus, to make it become more useful to the community.

\bibliography{bibliography}
\bibliographystyle{aaai}
\end{document}

%% file: LoZ.pdf_tex
%% Creator: Inkscape 0.91_64bit, www.inkscape.org
%% Accompanies image file 'LoZ.pdf' (pdf, eps, ps)
\begingroup%
  \makeatletter%
  \providecommand\color[2][]{%
    \errmessage{(Inkscape) Color is used for the text in Inkscape, but the package 'color.sty' is not loaded}%
    \renewcommand\color[2][]{}%
  }%
  \providecommand\transparent[1]{%
    \errmessage{(Inkscape) Transparency is used (non-zero) for the text in Inkscape, but the package 'transparent.sty' is not loaded}%
    \renewcommand\transparent[1]{}%
  }%
  \providecommand\rotatebox[2]{#2}%
  \ifx\svgwidth\undefined%
    \setlength{\unitlength}{595.27559055bp}%
    \ifx\svgscale\undefined%
      \relax%
    \else%
      \setlength{\unitlength}{\unitlength * \real{\svgscale}}%
    \fi%
  \else%
    \setlength{\unitlength}{\svgwidth}%
  \fi%
  \global\let\svgwidth\undefined%
  \global\let\svgscale\undefined%
  \makeatother%
  \begin{picture}(1,1.41428571)%
    \put(0,0){\includegraphics[width=\unitlength,page=1]{LoZ.pdf}}%
  \end{picture}%
\endgroup%

%% file: LoZSVG.pdf_tex
%% Creator: Inkscape 0.91_64bit, www.inkscape.org
%% Accompanies image file 'LoZSVG.pdf' (pdf, eps, ps)
\begingroup%
  \makeatletter%
  \providecommand\color[2][]{%
    \errmessage{(Inkscape) Color is used for the text in Inkscape, but the package 'color.sty' is not loaded}%
    \renewcommand\color[2][]{}%
  }%
  \providecommand\transparent[1]{%
    \errmessage{(Inkscape) Transparency is used (non-zero) for the text in Inkscape, but the package 'transparent.sty' is not loaded}%
    \renewcommand\transparent[1]{}%
  }%
  \providecommand\rotatebox[2]{#2}%
  \ifx\svgwidth\undefined%
    \setlength{\unitlength}{595.27559055bp}%
    \ifx\svgscale\undefined%
      \relax%
    \else%
      \setlength{\unitlength}{\unitlength * \real{\svgscale}}%
    \fi%
  \else%
    \setlength{\unitlength}{\svgwidth}%
  \fi%
  \global\let\svgwidth\undefined%
  \global\let\svgscale\undefined%
  \makeatother%
  \begin{picture}(1,1.41428571)%
    \put(0,0){\includegraphics[width=\unitlength,page=1]{LoZSVG.pdf}}%
    \put(0.30874056,1.12291328){\makebox(0,0)[b]{\smash{e}}}%
    \put(0,0){\includegraphics[width=\unitlength,page=2]{LoZSVG.pdf}}%
    \put(0.30034109,1.19598867){\makebox(0,0)[b]{\smash{b}}}%
    \put(0,0){\includegraphics[width=\unitlength,page=3]{LoZSVG.pdf}}%
    \put(0.46497072,0.85413021){\makebox(0,0)[b]{\smash{e,I}}}%
    \put(0,0){\includegraphics[width=\unitlength,page=4]{LoZSVG.pdf}}%
    \put(0.40785432,0.70797941){\makebox(0,0)[b]{\smash{e,i}}}%
    \put(0,0){\includegraphics[width=\unitlength,page=5]{LoZSVG.pdf}}%
    \put(0.44817178,0.78105481){\makebox(0,0)[b]{\smash{b}}}%
    \put(0,0){\includegraphics[width=\unitlength,page=6]{LoZSVG.pdf}}%
    \put(0.45825114,1.000281){\makebox(0,0)[b]{\smash{e,i}}}%
    \put(0,0){\includegraphics[width=\unitlength,page=7]{LoZSVG.pdf}}%
    \put(0.44901173,0.9272056){\makebox(0,0)[b]{\smash{k}}}%
    \put(0,0){\includegraphics[width=\unitlength,page=8]{LoZSVG.pdf}}%
    \put(0.5288067,0.70797941){\makebox(0,0)[b]{\smash{e,k}}}%
    \put(0,0){\includegraphics[width=\unitlength,page=9]{LoZSVG.pdf}}%
    \put(0.64975908,0.41567785){\makebox(0,0)[b]{\smash{I}}}%
    \put(0,0){\includegraphics[width=\unitlength,page=10]{LoZSVG.pdf}}%
    \put(0.65647866,0.56182862){\makebox(0,0)[b]{\smash{p}}}%
    \put(0,0){\includegraphics[width=\unitlength,page=11]{LoZSVG.pdf}}%
    \put(0.6430395,0.48875325){\makebox(0,0)[b]{\smash{l}}}%
    \put(0,0){\includegraphics[width=\unitlength,page=12]{LoZSVG.pdf}}%
    \put(0.40281464,0.56182862){\makebox(0,0)[b]{\smash{e,k}}}%
    \put(0,0){\includegraphics[width=\unitlength,page=13]{LoZSVG.pdf}}%
    \put(0.40281464,0.41567785){\makebox(0,0)[b]{\smash{e}}}%
    \put(0,0){\includegraphics[width=\unitlength,page=14]{LoZSVG.pdf}}%
    \put(0.26674321,0.41567785){\makebox(0,0)[b]{\smash{e}}}%
    \put(0,0){\includegraphics[width=\unitlength,page=15]{LoZSVG.pdf}}%
    \put(0.32217971,0.48875325){\makebox(0,0)[b]{\smash{l}}}%
    \put(0,0){\includegraphics[width=\unitlength,page=16]{LoZSVG.pdf}}%
    \put(0.36921675,0.78105481){\makebox(0,0)[b]{\smash{b}}}%
    \put(0,0){\includegraphics[width=\unitlength,page=17]{LoZSVG.pdf}}%
    \put(0.39021543,0.34260246){\makebox(0,0)[b]{\smash{k}}}%
    \put(0,0){\includegraphics[width=\unitlength,page=18]{LoZSVG.pdf}}%
    \put(0.28186225,0.14689478){\makebox(0,0)[b]{\smash{e,k}}}%
    \put(0,0){\includegraphics[width=\unitlength,page=19]{LoZSVG.pdf}}%
    \put(0.40281464,0.14689478){\makebox(0,0)[b]{\smash{e,k}}}%
    \put(0,0){\includegraphics[width=\unitlength,page=20]{LoZSVG.pdf}}%
    \put(0.52376702,0.14689478){\makebox(0,0)[b]{\smash{s}}}%
    \put(0,0){\includegraphics[width=\unitlength,page=21]{LoZSVG.pdf}}%
    \put(0.40869427,0.34260246){\makebox(0,0)[b]{\smash{k}}}%
    \put(0,0){\includegraphics[width=\unitlength,page=22]{LoZSVG.pdf}}%
    \put(0.41793368,0.78105481){\makebox(0,0)[b]{\smash{b}}}%
    \put(0,0){\includegraphics[width=\unitlength,page=23]{LoZSVG.pdf}}%
    \put(0.25750379,0.78105481){\makebox(0,0)[b]{\smash{k}}}%
    \put(0,0){\includegraphics[width=\unitlength,page=24]{LoZSVG.pdf}}%
    \put(0.52376702,0.41567785){\makebox(0,0)[b]{\smash{t}}}%
    \put(0,0){\includegraphics[width=\unitlength,page=25]{LoZSVG.pdf}}%
    \put(0.53384638,0.56182862){\makebox(0,0)[b]{\smash{b}}}%
    \put(0,0){\includegraphics[width=\unitlength,page=26]{LoZSVG.pdf}}%
    \put(0.53384638,0.48875325){\makebox(0,0)[b]{\smash{l}}}%
    \put(0,0){\includegraphics[width=\unitlength,page=27]{LoZSVG.pdf}}%
    \put(0.58928289,0.85413021){\makebox(0,0)[b]{\smash{e,k,p}}}%
    \put(0,0){\includegraphics[width=\unitlength,page=28]{LoZSVG.pdf}}%
    \put(0.64975908,0.70797941){\makebox(0,0)[b]{\smash{e,k}}}%
    \put(0,0){\includegraphics[width=\unitlength,page=29]{LoZSVG.pdf}}%
    \put(0.6304403,0.78105481){\makebox(0,0)[b]{\smash{k}}}%
    \put(0,0){\includegraphics[width=\unitlength,page=30]{LoZSVG.pdf}}%
    \put(0.46917046,0.9272056){\makebox(0,0)[b]{\smash{k}}}%
    \put(0,0){\includegraphics[width=\unitlength,page=31]{LoZSVG.pdf}}%
    \put(0.72703421,0.78105481){\makebox(0,0)[b]{\smash{b}}}%
    \put(0,0){\includegraphics[width=\unitlength,page=32]{LoZSVG.pdf}}%
    \put(0.19366781,0.78105481){\makebox(0,0)[b]{\smash{k}}}%
    \put(0,0){\includegraphics[width=\unitlength,page=33]{LoZSVG.pdf}}%
    \put(0.54728553,0.63490402){\makebox(0,0)[b]{\smash{l}}}%
    \put(0,0){\includegraphics[width=\unitlength,page=34]{LoZSVG.pdf}}%
    \put(0.60356199,0.78105481){\makebox(0,0)[b]{\smash{k}}}%
    \put(0,0){\includegraphics[width=\unitlength,page=35]{LoZSVG.pdf}}%
    \put(0.66067839,0.63490402){\makebox(0,0)[b]{\smash{k}}}%
    \put(0,0){\includegraphics[width=\unitlength,page=36]{LoZSVG.pdf}}%
    \put(0.53804612,0.63490402){\makebox(0,0)[b]{\smash{k}}}%
    \put(0,0){\includegraphics[width=\unitlength,page=37]{LoZSVG.pdf}}%
    \put(0.65815855,0.48875325){\makebox(0,0)[b]{\smash{l}}}%
    \put(0,0){\includegraphics[width=\unitlength,page=38]{LoZSVG.pdf}}%
    \put(0.64891913,0.63490402){\makebox(0,0)[b]{\smash{k}}}%
  \end{picture}%
\endgroup%

%% file: E1M1.pdf_tex
%% Creator: Inkscape 0.91_64bit, www.inkscape.org
%% Accompanies image file 'E1M1.pdf' (pdf, eps, ps)
\begingroup%
  \makeatletter%
  \providecommand\color[2][]{%
    \errmessage{(Inkscape) Color is used for the text in Inkscape, but the package 'color.sty' is not loaded}%
    \renewcommand\color[2][]{}%
  }%
  \providecommand\transparent[1]{%
    \errmessage{(Inkscape) Transparency is used (non-zero) for the text in Inkscape, but the package 'transparent.sty' is not loaded}%
    \renewcommand\transparent[1]{}%
  }%
  \providecommand\rotatebox[2]{#2}%
  \ifx\svgwidth\undefined%
    \setlength{\unitlength}{819.2bp}%
    \ifx\svgscale\undefined%
      \relax%
    \else%
      \setlength{\unitlength}{\unitlength * \real{\svgscale}}%
    \fi%
  \else%
    \setlength{\unitlength}{\svgwidth}%
  \fi%
  \global\let\svgwidth\undefined%
  \global\let\svgscale\undefined%
  \makeatother%
  \begin{picture}(1,0.61621094)%
    \put(0,0){\includegraphics[width=\unitlength,page=1]{E1M1.pdf}}%
  \end{picture}%
\endgroup%

%% file: main.bbl
\begin{thebibliography}{}

\bibitem[\protect\citeauthoryear{{ Sierra On-Line}}{1987}]{KINGS_QUEST}
{ Sierra On-Line}.
\newblock 1987.
\newblock {King's Quest}.

\bibitem[\protect\citeauthoryear{Dahlskog and Togelius}{2012}]{PATTERNS_PCG}
Dahlskog, S., and Togelius, J.
\newblock 2012.
\newblock Patterns and procedural content generation: Revisiting mario in world
  1 level 1.
\newblock In {\em Proceedings of the First Workshop on Design Patterns in
  Games}.

\bibitem[\protect\citeauthoryear{Dahlskog, Togelius, and
  Nelson}{2014}]{DAHLSKOG}
Dahlskog, S.; Togelius, J.; and Nelson, M.~J.
\newblock 2014.
\newblock Linear levels through n-grams.
\newblock In {\em Proceedings of the 18th International Academic MindTrek
  Conference}.

\bibitem[\protect\citeauthoryear{Davies}{1990}]{COCA}
Davies, M.
\newblock 1990.
\newblock {The Corpus of Contemporary American English: 520 million words}.

\bibitem[\protect\citeauthoryear{Dormans}{2010}]{DORMANS_ZELDA}
Dormans, J.
\newblock 2010.
\newblock Adventures in level design.
\newblock In {\em Workshop on PCG in Games}.

\bibitem[\protect\citeauthoryear{Gansner and North}{2000}]{GRAPHVIZ}
Gansner, E.~R., and North, S.~C.
\newblock 2000.
\newblock An open graph visualization system and its applications to software
  engineering.
\newblock {\em SOFTWARE - PRACTICE AND EXPERIENCE}.

\bibitem[\protect\citeauthoryear{Gatys, Ecker, and Bethge}{2015}]{NEURAL_STYLE}
Gatys, L.~A.; Ecker, A.~S.; and Bethge, M.
\newblock 2015.
\newblock A neural algorithm of artistic style.
\newblock {\em CoRR} abs/1508.06576.

\bibitem[\protect\citeauthoryear{Godfrey, Holliman, and
  McDaniel}{1992}]{SWITCHBOARD}
Godfrey, J.~J.; Holliman, E.~C.; and McDaniel, J.
\newblock 1992.
\newblock Switchboard: telephone speech corpus for research and development.
\newblock In {\em Acoustics, Speech, and Signal Processing}.

\bibitem[\protect\citeauthoryear{Guzdial and Riedl}{2015}]{GUZDIAL}
Guzdial, M., and Riedl, M.~O.
\newblock 2015.
\newblock Toward game level generation from gameplay videos.
\newblock In {\em Proceedings of the FDG workshop on PCG in Games}.

\bibitem[\protect\citeauthoryear{Hoover, Togelius, and
  Yannakakis}{2015}]{HOOVER}
Hoover, A.~K.; Togelius, J.; and Yannakakis, G.~N.
\newblock 2015.
\newblock Composing video game levels with music metaphors through functional
  scaffolding.
\newblock In {\em ICCC Workshop on CCG}.

\bibitem[\protect\citeauthoryear{{id Software}}{1993}]{DOOM}
{id Software}.
\newblock 1993.
\newblock {Doom}.

\bibitem[\protect\citeauthoryear{{id Software}}{1994}]{DOOM2}
{id Software}.
\newblock 1994.
\newblock {Doom II: Hell on Earth}.

\bibitem[\protect\citeauthoryear{{Infocom}}{1987}]{ZORK}
{Infocom}.
\newblock 1987.
\newblock {Zork: The Great Underground Empire - Part I }.

\bibitem[\protect\citeauthoryear{Krizhevsky, Nair, and Hinton}{2016}]{CIFAR}
Krizhevsky, A.; Nair, V.; and Hinton, G.
\newblock 2016.
\newblock {CIFAR-10 and CIFAR-100}.

\bibitem[\protect\citeauthoryear{Lecun and Cortes}{2016}]{MNIST}
Lecun, Y., and Cortes, C.
\newblock 2016.
\newblock {The MNIST database of handwritten digits}.

\bibitem[\protect\citeauthoryear{{Nintendo EAD }}{1991}]{LTTP}
{Nintendo EAD }.
\newblock 1991.
\newblock {The Legend of Zelda: A Link to the Past}.

\bibitem[\protect\citeauthoryear{{Nintendo EAD }}{1992}]{MARIO_KART}
{Nintendo EAD }.
\newblock 1992.
\newblock {Super Mario Kart}.

\bibitem[\protect\citeauthoryear{{Nintendo EAD }}{1993}]{LINKS_AWAKENING}
{Nintendo EAD }.
\newblock 1993.
\newblock {The Legend of Zelda: Link's Awakening}.

\bibitem[\protect\citeauthoryear{{Nintendo RD1}}{1983}]{SML}
{Nintendo RD1}.
\newblock 1983.
\newblock {Super Mario Land}.

\bibitem[\protect\citeauthoryear{{Nintendo RD1}}{1986}]{KID_ICARUS}
{Nintendo RD1}.
\newblock 1986.
\newblock {Kid Icarus}.

\bibitem[\protect\citeauthoryear{{Nintendo RD4 }}{1986}]{LOZ}
{Nintendo RD4 }.
\newblock 1986.
\newblock {The Legend of Zelda}.

\bibitem[\protect\citeauthoryear{{Nintendo RD4}}{1983}]{SUPER_MARIO_BROS}
{Nintendo RD4}.
\newblock 1983.
\newblock {Super Mario Brothers}.

\bibitem[\protect\citeauthoryear{{n}o and Missura}{2015}]{GRAPH_GRAMMAR_SMB}
{n}o, S.~L., and Missura, O.
\newblock 2015.
\newblock Graph grammars for super mario bros levels.
\newblock In {\em FDG workshop on PCG in Games}.

\bibitem[\protect\citeauthoryear{Paul and Baker}{1992}]{WSJ}
Paul, D.~B., and Baker, J.~M.
\newblock 1992.
\newblock The design for the wall street journal-based csr corpus.

\bibitem[\protect\citeauthoryear{Shaker and Abou-Zleikha}{2014}]{SHAKER}
Shaker, N., and Abou-Zleikha, M.
\newblock 2014.
\newblock Alone we can do so little, together we can do so much.
\newblock In {\em AIIDE}.

\bibitem[\protect\citeauthoryear{Smith \bgroup et al\mbox.\egroup
  }{2012}]{REFRACTION}
Smith, A.~M.; Andersen, E.; Mateas, M.; and Popovi{\'c}, Z.
\newblock 2012.
\newblock A case study of expressively constrainable level design automation
  tools for a puzzle game.
\newblock In {\em FDG}.

\bibitem[\protect\citeauthoryear{Smith, Whitehead, and Mateas}{2010}]{TANAGRA}
Smith, G.; Whitehead, J.; and Mateas, M.
\newblock 2010.
\newblock Tanagra: a mixed-initiative level design tool.
\newblock In {\em FDG}.

\bibitem[\protect\citeauthoryear{Smith}{1983}]{LODE_RUNNER}
Smith, D.
\newblock 1983.
\newblock {Lode Runner}.

\bibitem[\protect\citeauthoryear{Snodgrass and {n}\'{o}n}{2013}]{SNODGRASS}
Snodgrass, S., and {n}\'{o}n, S.~O.
\newblock 2013.
\newblock Generating maps using markov chains.
\newblock In {\em AIIDE}.

\bibitem[\protect\citeauthoryear{Snodgrass and {n}\'{o}n}{2014}]{SNODGRASS2}
Snodgrass, S., and {n}\'{o}n, S.~O.
\newblock 2014.
\newblock Experiments in map generation using markov chains.
\newblock In {\em FDG}.

\bibitem[\protect\citeauthoryear{Sorenson and Pasquier}{2010}]{SORENSON}
Sorenson, N., and Pasquier, P.
\newblock 2010.
\newblock Towards a generic framework for automated video game level creation.
\newblock In {\em Applications of Evolutionary Computation}.

\bibitem[\protect\citeauthoryear{Summerville and
  Mateas}{2015}]{SAMPLING_HYRULE}
Summerville, A.~J., and Mateas, M.
\newblock 2015.
\newblock Sampling hyrule: Multi-technique probabilistic level generation for
  action role playing games.
\newblock In {\em AIIDE}.

\bibitem[\protect\citeauthoryear{Summerville and Mateas}{2016}]{SMBRNN}
Summerville, A., and Mateas, M.
\newblock 2016.
\newblock Super mario as a string: Platformer level generation via lstms.

\bibitem[\protect\citeauthoryear{Summerville \bgroup et al\mbox.\egroup
  }{2015}]{BAYESZELDA}
Summerville, A.; Behrooz, M.; Mateas, M.; and Jhala, A.
\newblock 2015.
\newblock The learning of zelda: Data-driven learning of level topology.
\newblock In {\em FDG}.

\bibitem[\protect\citeauthoryear{Summerville, Philip, and
  Mateas}{2015}]{MCMCTS}
Summerville, A.; Philip, S.; and Mateas, M.
\newblock 2015.
\newblock Mcmcts pcg 4 smb: Monte carlo tree search to guide platformer level
  generation.
\newblock In {\em AIIDE}.

\bibitem[\protect\citeauthoryear{{Taito}}{1987}]{RAINBOW_ISLANDS}
{Taito}.
\newblock 1987.
\newblock {Rainbow Islands: The Story of Bubble Bobble 2}.

\bibitem[\protect\citeauthoryear{Toy, Wichman, and Arnold}{1980}]{ROGUE}
Toy, M.; Wichman, G.; and Arnold, K.
\newblock 1980.
\newblock {Rogue}.

\end{thebibliography}
